\documentclass[journal]{new-aiaa}
\usepackage[utf8]{inputenc}
\usepackage{float}
\usepackage{graphicx}
\usepackage{amsmath}
\usepackage[version=4]{mhchem}
\usepackage{siunitx}
\usepackage{longtable,tabularx}
\usepackage{subcaption}
\usepackage{multirow}
\usepackage{adjustbox}
\usepackage{bm}
\usepackage{gensymb}

\let\svthefootnote\thefootnote
\newcommand\freefootnote[1]{
  \let\thefootnote\relax
  \footnotetext{#1}
  \let\thefootnote\svthefootnote
}

\newcommand{\rev}[1]{\textcolor{black}{#1}}

\title{Space Logistics Analysis and Incentive Design for
Commercialization of Orbital Debris Remediation}

\author{Asaad Abdul-Hamid\footnote{Ph.D. Candidate, Department of Systems and Enterprises, AIAA Student Member.}} 
\affil{Stevens Institute of Technology, Hoboken, NJ, 07030}
\author{
Brycen D. Pearl\footnote{Ph.D. Student, Department of Mechanical, Materials and Aerospace Engineering, AIAA Student Member.} and Hang Woon Lee\footnote{Assistant Professor, Department of Mechanical, Materials and Aerospace Engineering. AIAA Member.}
}
\affil{West Virginia University, Morgantown, WV, 26506}
\author{Hao Chen\footnote{Assistant Professor, Department of Systems and Enterprises; hao.chen@stevens.edu. AIAA Member. (Corresponding Author)}} 
\affil{Stevens Institute of Technology, Hoboken, NJ, 07030}

\begin{document}

\newpage 

\freefootnote{This paper is a substantially revised version of the paper AIAA 2025-1105, presented at the 2025 AIAA SciTech Forum, Orlando, FL, January 6-10, 2025.}

\maketitle

\begin{abstract}
As orbital debris continues to become a higher priority for the space industry, there is a need to explore how partnerships between the public and private space sector may aid in addressing this issue. This research develops a space logistics framework for planning orbital debris remediation missions, providing a quantitative basis for partnerships that are mutually beneficial between space operators and debris remediators. By integrating network-based space logistics and game theory, we illuminate the high-level costs of remediating orbital debris, and the surplus that stands to be shared as a result. These findings indicate significant progress toward the continued development of a safe, sustainable, and profitable space economy.
\end{abstract}

\section*{Nomenclature}

{\renewcommand\arraystretch{1.0}
\noindent\begin{longtable*}{@{}l @{\quad=\quad} l@{}}
$\mathcal{A}$  & set of arcs \\
$\mathcal{B}$  & remediation mission benefit \\
$c$ &    cost coefficient \\
$\mathcal{D}$   & set of debris\\
$d$& mission demand \\
\(\mathit{e}_v\)& spacecraft design parameters\\
${F}$  & remediation mission fee \\
${G}$  & time-expanded network \\
\(\mathit{f}_v\)& spacecraft fuel type\\
$H$ & concurrency constraint matrix\\
$I$ & number of complete satellite revolutions\\ 
$\mathcal{J}$   & remediation mission cost\\
$J$ & number of complete satellite revolutions\\
$M$ & mean anomaly\\
$\mathcal{N}$ & set of nodes \\
$n$ & mean motion\\
$P_{EA}$ & aggregate probability of encounter\\
$\overline{P}_{Ej}$ & no-encounter probability of $j$th pair\\
$Q$ & commodity transformation matrix \\
${R}$  & remediation mission incentive \\
$r$ & disagreement point utility \\
\(\mathit{s}_v\)& structure mass of spacecraft \(\mathit{v}\)\\
$\mathcal{T}$ & set of time steps \\
$\mathcal{U}$ & total utility \\
$\mathcal{V}$ & set of spacecraft \\
$W$   & set of time windows \\
$x$  & commodity flow variable \\
$y$  & commodity flow variable \\
$\mathcal{Z}$ & Nash product \\
$\Delta v$ & change of velocity, km/s\\
$\epsilon$ & resonance parameter\\
$\theta$  & incentive coefficient \\
\multicolumn{2}{l}{\textit{Subscripts}}\\
$i$ & node index\\
$j$ & node index\\
$l$ & concurrency index\\
$p$ & commodity index\\
$t$ & time step index\\
$v$ & spacecraft index\\
\end{longtable*}}

\section{Introduction}
\lettrine{P}{rivate} sector participation in space has been largely beneficial to space development, but brings a more complex network of stakeholders along with the larger pool of funding towards space projects. This is in contrast to the previous space age, which was dominated by government entities, and constrained by the resources and operational capabilities that those entities could support. In this new age of space commercialization, private companies reduce financial pressures and share the risk of space projects with public space agencies. Healthy public-private partnerships generate value for countries by increasing agility and competitiveness within the particular industry. Establishing these beneficial relationships can only be done through the implementation of adequate incentive mechanisms, which assist private companies in overcoming the initial barriers to entering the market. This has been seen with Airbus in Europe~\cite{Neven_Seabright_Grossman_1995}, the photovoltaic industry in China \cite{Zhao_Zhang_Hubbard_Yao_2013}, and the shale gas industry in the United States \cite{Whitton_Brasier_Charnley-Parry_Cotton_2017}. Within the space industry, NASA initiated the Commercial Orbital Transportation Services program in 2006. This awarded SpaceX and Rocketplane Kistler partial funding to develop U.S. launch capacities for resupplying the International Space Station \cite{NASA_2014}. These programs were all developed with strong incentive mechanisms that eventually resulted in significant social and economic returns. Developing this type of robust partnership is crucial to addressing the increasingly growing problem of orbital debris.  
Orbital debris is a rising priority within the space industry. Rocket stages, defunct satellites, fragmentation debris, and other such detritus comprise a population of dangerous objects that orbit Earth at multiple kilometers per second, posing a significant threat to any operational spacecraft in their path. The European Space Agency (ESA) estimates that there are over 34,000 large debris objects in Earth’s orbit, with the majority of debris objects being concentrated in Low Earth Orbit (LEO) \cite{ESA_2024}. The world’s first active debris remediation mission is scheduled for 2026 \cite{ESA_2024a}, after which it can be expected that even more active debris remediation missions will be planned and executed. 

Current literature for active debris remediation mission architecture does not explore operational details. Many scholars and professionals, such as Colvin et al \cite{Colvin_2023}, have given estimates for the total cost of debris remediation missions based on a high-level concept of operations (CONOPS) of various remediation strategies. The overall cost for the same remediation method can vary greatly from study to study. As an example, the cost estimation for controlled debris re-entry varies from \$4,000-\$60,000/kg \cite{Colvin_2023, OlearyConf,Estable}, and the cost estimation for uncontrolled reentry varies from \$3,000-\$40,000/kg \cite{Colvin_2023,Duchek_Abrams_Infeld_Jolly_Drews_Hopkins_2015,Rainbow_2023}. While recent literature has reported that there can be enough near-term benefits to passively incentivize orbital debris remediation, it does not explore proactive incentive mechanisms that would be needed to stimulate private sector engagement in active debris remediation operations \cite{Colvin_2023, OlearyConf, Liou_2011}. There is a need for economic evaluations that consider the details of active debris remediation operations, and assess the incentive mechanisms necessary for commercial participation. 

\begin{figure}[h]
    \centering
    \includegraphics[width=\textwidth]{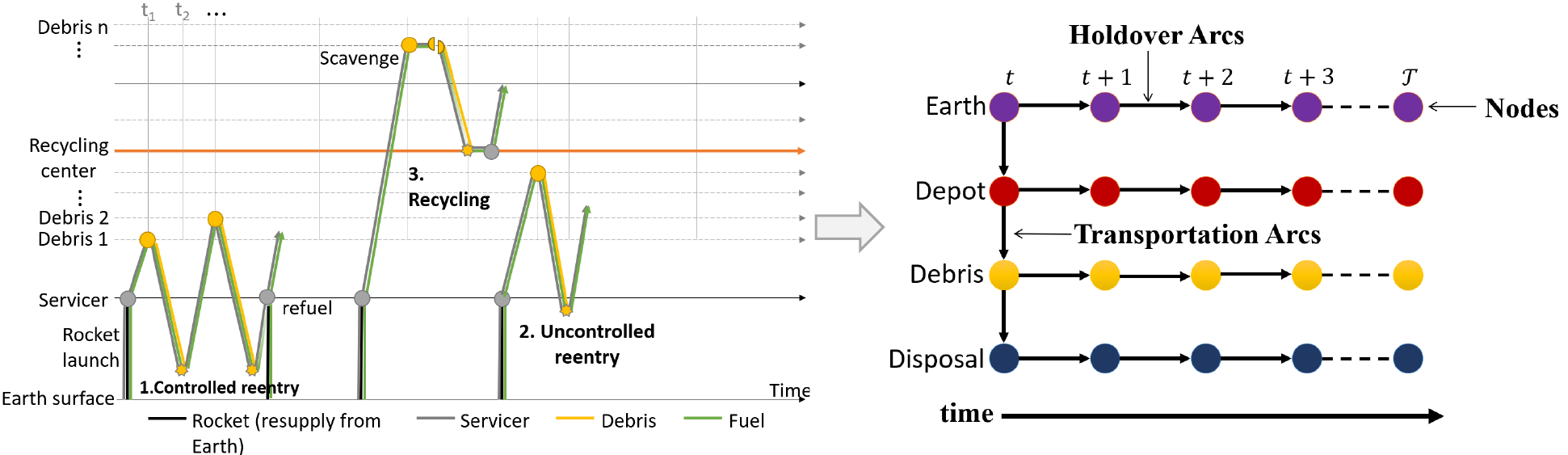}
    \caption{\rev{Logistics Network Overview}}
    \label{fig:networkoverview}
\end{figure}

We propose a design and analysis framework which integrates a network-based space logistics model and a game theory-based incentive design model. This framework will yield comprehensive assessments of mission structure and incentive mechanisms for different orbital debris remediation methods. We shall consider three remediation methods: controlled reentry, uncontrolled reentry, and recycling. As a simulation, we will consider the 50 statistically most concerning large debris \cite{McKnight2021} as the targets for remediation. Our cost and benefit estimations can be traced back to a specific time step thanks to our space logistics model. This enables typical capital budgeting techniques like payback period, net present value, and internal rate of return \cite{Alesii_2006,Rushinck_1983}, which need time-based cost and benefit information to support commercial decision-making. Our logistics model evolves the generalized multi-commodity network flow framework \cite{Sarton,Chen_Ornik_Ho_2022,Chen_Ho_2018,Ishimatsu_logistics0_2016,HO_logistics_201651}, and has been tailored specifically for active debris remediation methods. The operational details unique to each remediation method are accounted for as essential decision variables at each appropriate time step. Figure \ref{fig:networkoverview} shows an overview of the logistics network for debris remediation.

The remainder of this paper is structured as follows. Section
\ref{Methodology} introduces the methodology for our space logistics cost model, the probability-based benefit model, and the game theory model for incentive design. Section \ref{Results and Analysis} evaluates the framework by considering a mission structure to remove the top 50 most concerning objects in LEO, presenting and discussing the results for each model in the framework. Section \ref{Conclusion} concludes the paper and offers suggestions on how the work might evolve in the future.

\section{Methodology}\label{Methodology}
\subsection{Benefit Evaluation}

To evaluate the benefit of remediating debris, we consider the benefit to be a function of the cost of warnings, maneuvers, and collisions that can be avoided in the absence of debris. This applies to the large debris we are directly removing, and all of the small debris that the large debris may generate via catastrophic collisions. As such, this section covers the methodology for calculating total benefit as a function of both small and large debris.

Given the inherent difficulty in tracking small debris, our methodology relies on modeling techniques to estimate their distribution. Innovative tracking methods, such as the Goldstone Orbital Debris Radar operated by NASA’s Jet Propulsion Laboratory, have enhanced our capacity to monitor debris as diminutive as 2-3 mm at altitudes below 1000 km \cite{Manis}.

For our analysis, we utilize the model developed by the Comprehensive Space Operations Center (COMSPOC), specifically their Number of Encounters Assessment Tool (NEAT), to assess collision risks and evaluate the potential benefits of debris removal for satellites. NEAT allows users to customize parameters such as constellation size, orbit altitude, and inclination, while also defining thresholds for warnings, maneuvers, and collisions. This open-source, web-based tool facilitates precise calculations of warning occurrences, required maneuvers, and potential collisions over user-defined timeframes \cite{Alfano_Oltrogge_2018}.

\rev{In assessing the frequency of satellite conjunction events, Alfano and Oltrogge \cite{Alfano_Oltrogge_2018} developed a volumetric approach that quantifies satellite encounter rates within a specified orbital regime. This volumetric approach is the mathematical foundation of NEAT.} The methodology is particularly useful for characterizing the likelihood of close approaches between satellites, which are critical for operational planning and collision avoidance strategies.

The methodology defines an encounter volume as an ellipsoid in the Radial-InTrack-CrossTrack (RIC) frame centered on the secondary satellite. This ellipsoid remains constant in size, shape, and orientation relative to the satellite's orbital path. The dimensions of the ellipsoid are significantly larger than the satellites themselves, allowing the model to focus on the volumetric region where potential encounters might occur rather than on direct collision probabilities.
For a given pair of satellites, the primary satellite's orbit is evaluated to determine whether it intersects the encounter volume surrounding the secondary satellite. This is achieved by incrementally advancing the ellipsoid along the secondary satellite’s orbit and checking for any overlap with the primary satellite's trajectory.

The probability of an encounter \(\mathit{P_E}\) is computed by summing the probabilities over small increments of mean anomaly $M$, representing the satellites' positions along their orbits. Specifically, \(\mathit{P_E}\) is given by:

\begin{equation}\label{probencount}
    P_E= \sum_{i} \left(\frac{\Delta M_{i1}}{2\pi}\right)\left(\frac{\Delta M_{i2}}{2\pi}\right)
\end{equation}

\noindent where \(\Delta M_{i1}\) and \(\Delta M_{i2}\) are the ranges of mean anomaly during which the primary and secondary satellites are within the encounter volume. This summation accounts for the different potential intersection points along the satellites' orbits.

For multiple satellite pairs, the aggregate probability \(P_{EA}\) that at least one encounter occurs is determined using:

\begin{equation}\label{aggprob}
P_{EA}=  1- \prod_{j}(\overline{P}_{Ej})
\end{equation}

\noindent where \(\overline{P}_{Ej}\) is the no-encounter probability for the \(\mathit{j}\)th pair. This approach provides a comprehensive estimate of the encounter frequency over a specified period.

The methodology also considers orbital resonances, where satellites revisit each other’s orbital positions periodically. In such cases, the assumption of uniformly distributed mean anomaly does not hold. The resonance condition is identified when the difference in mean motion \(\mathit{n}_1\) and \(\mathit{n}_2\) between the two satellites satisfies:

\begin{equation}\label{rescondi}
    |J\cdot n_1 - I \cdot n_2| \leq \epsilon
\end{equation}

\noindent where \(\mathit{I}\) and \(\mathit{J}\) are integers representing the number of complete satellite revolutions, and $\epsilon$ is the resonance parameter.

To simulate the impact of debris removal on satellite operations, we configure NEAT with a single-satellite constellation over a one-year period, with a presumed space population of 14,450 objects based on ESA Space Environmental Statistics as of December 2023. Threshold parameters recommended by NEAT \cite{AlfanoConf} are adopted: warnings at 3000 m, maneuvers at 1000 m, and collision risk at 5 m. NEAT's approach to parameter assessment involves encapsulating the satellite within a spherical boundary defined by its longest axis. This innovative method enables a volumetric analysis, allowing for precise evaluation of the proximity between modeled debris and the simulated satellite constellation.

\begin{figure}[h]
    \centering
    \includegraphics[width=0.7\textwidth]{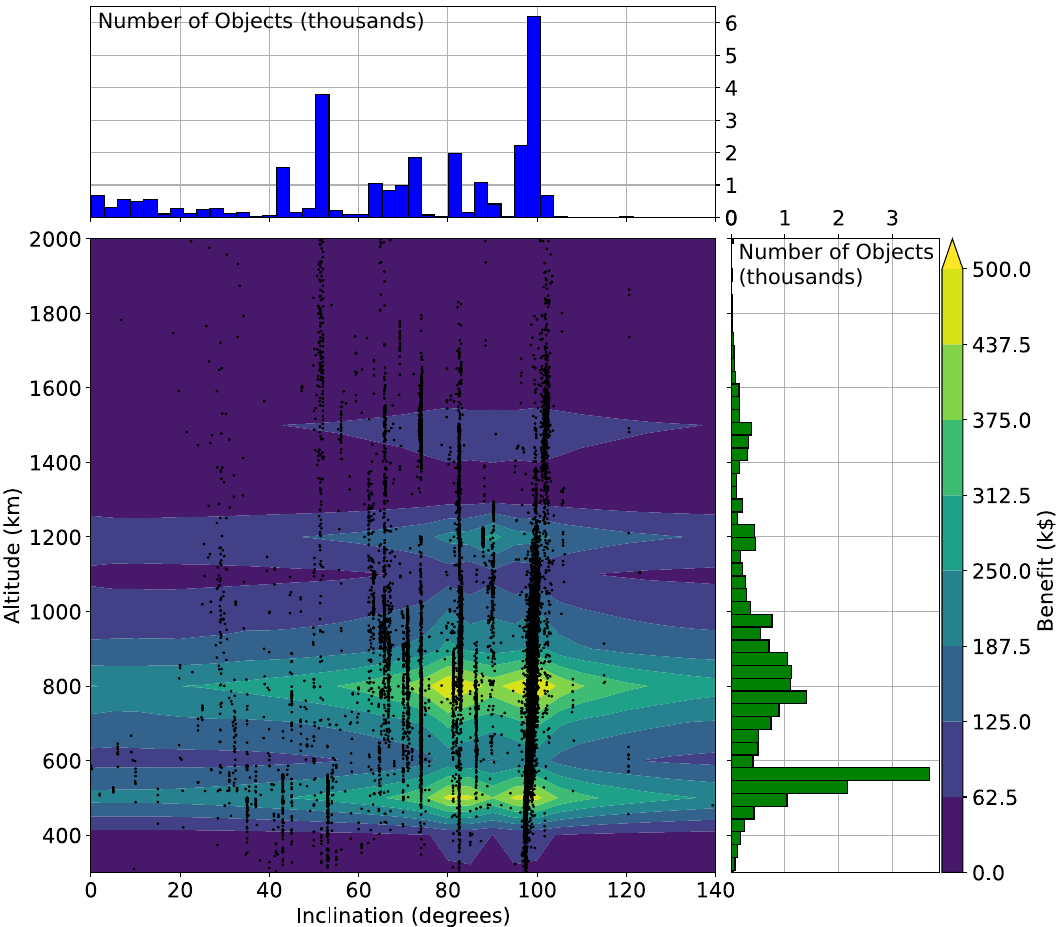}
    \caption{ \rev{Small Debris ($<$ 10 cm) Removal Benefit Contour Plot with Black Dots Representing In-Orbit Objects}}
    \label{fig:smallbenobject}
\end{figure}

Data collection involves assessing warnings, maneuvers, and collisions at varying altitudes within LEO (300 km to 2000 km) in 100 km increments, and inclinations ($0^\circ$ to $180^\circ$) in 5-degree increments. The cost framework proposed by Thomas et al. \cite{Colvin_2023} is applied: \$100 per warning, \$1000 per maneuver, and \$500 million per collision.
Multiplying these costs by the frequency of warnings, maneuvers, and collisions per year yields the perturbation cost for one debris object. We designate that the benefit of removing debris is equal to the avoided perturbation cost. Finally, to obtain a total yearly benefit (including all 3 perturbations) for removing each piece of debris, we can sum warning, maneuver, and collision costs. Figure \ref{fig:smallbenobject} displays a contour plot of the calculated benefit for small debris removal, with black dots representing current in-orbit objects. \rev{We define small debris objects as being debris with a characteristic length smaller than 10 cm.}

\begin{figure}[h]
    \centering
    \includegraphics[width=0.7\textwidth]{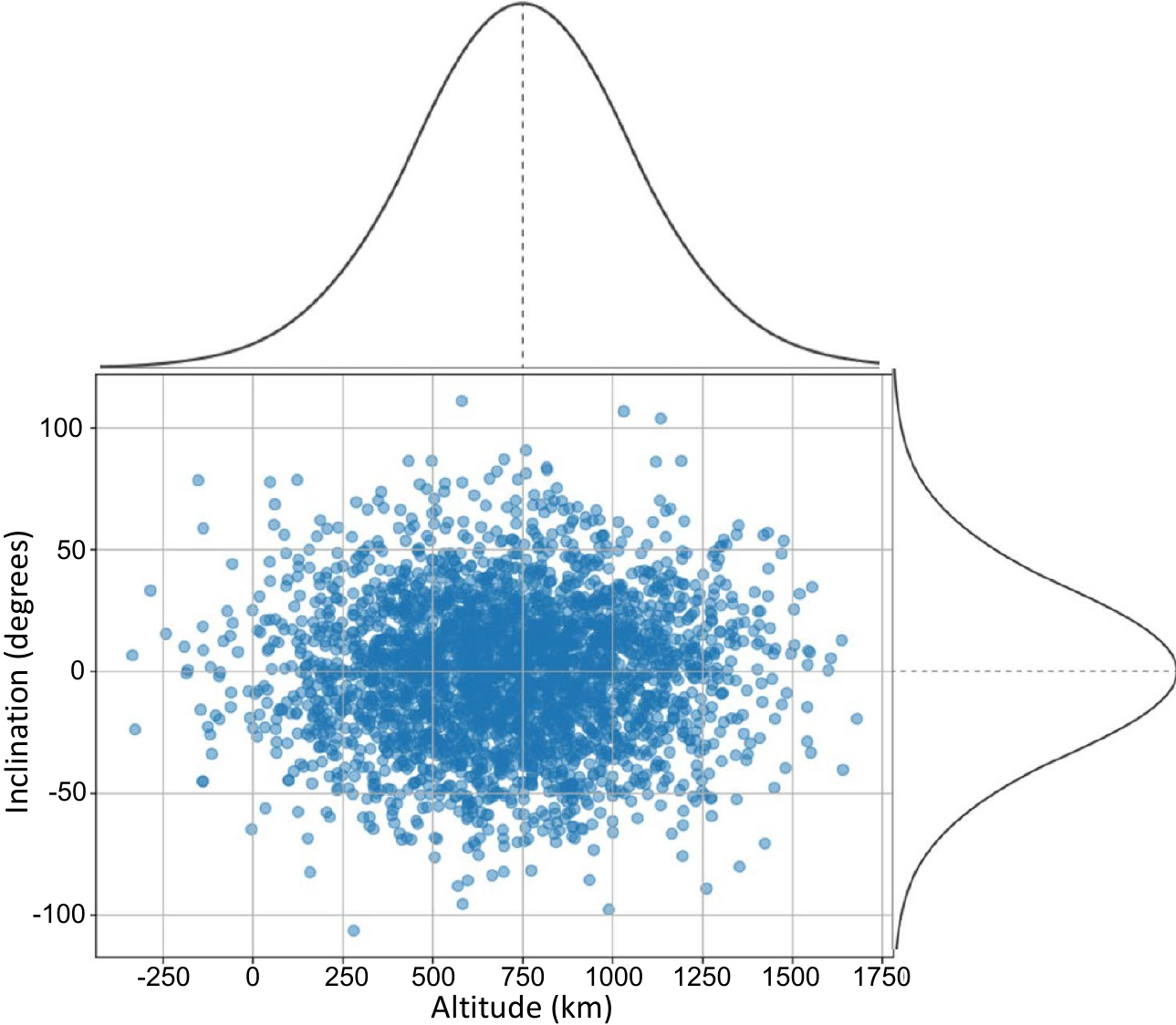}
    \caption{\rev{Normal Debris Distribution for Altitude and Inclination at 750 km and 0º}}
    \label{fig:normaldis}
\end{figure}

In this study, we aim to assess the benefits of removing large debris pieces that involve additional benefits because of their potential fragmentation into smaller fragments. To achieve this, the initial step involves examining the fragmentation process of large debris into smaller fragments. Drawing from the insights provided by Ref \cite{Colvin_2023}, which investigates scenarios such as the Anti-Satellite (ASAT) scenario, we estimate that approximately 3000 small fragments are generated following a collision with a large object.
The distribution of these newly formed debris fragments in space follows a normal distribution roughly for both altitude and inclination parameters \cite{Colvin_2023}.\rev{ We use this distribution to generate  Fig. \ref{fig:normaldis}, showing an example using a mean altitude of 750 km and a mean inclination of 0 degrees.}

To establish the double normal distribution encompassing both altitude and inclination, it is essential to determine the standard deviation for each parameter. For the altitude standard deviation, we refer to Ref \cite{Colvin_2023}, which discusses the distribution of ASAT debris across various altitudes. From this distribution, we derive a standard deviation of 313.74 km for altitude and $24.6^{\circ}$ for inclination.
These standard deviation values serve as essential parameters in our modeling framework, enabling us to simulate the spatial distribution of debris fragments resulting from the fragmentation of large debris objects.

\rev{After generating these normal distributions for 648 combinations of altitudes and inclinations, we consider the 3,000 debris fragments that would result from a catastrophic collision at any of those locations. This results in a dataset comprising 1,944,000 randomized values, ranging between 300 and 2000 km in altitude and from $0^{\circ}$ to $180^{\circ}$ in inclination. Using the probability of collisions from NEAT, we can calculate the additional benefit of preventing catastrophic collisions.} During the generation of these distributions, we encountered a challenge where some randomized values fell outside of the desired ranges. For values below $0^{\circ}$ in inclination, we simply added an additional $180^{\circ}$ to bring them within the valid range. For values outside the altitude range of interest, we set them to zero. Although this manipulation results in zero values for certain rows in the dataset, it does not impact the overall calculations of sums or differences. \rev{When assessing the benefit of removing debris, we need to consider the reduction in population and how long a given debris object would have remained in orbit. In our analysis, this is encapsulated by the debris-years metric, which is explored further in Section \ref{Results and Analysis}.}


\subsection{Space Logistic Model for Cost Evaluation}

\begin{figure}[h]
    \centering
    \includegraphics[width=.8\textwidth]{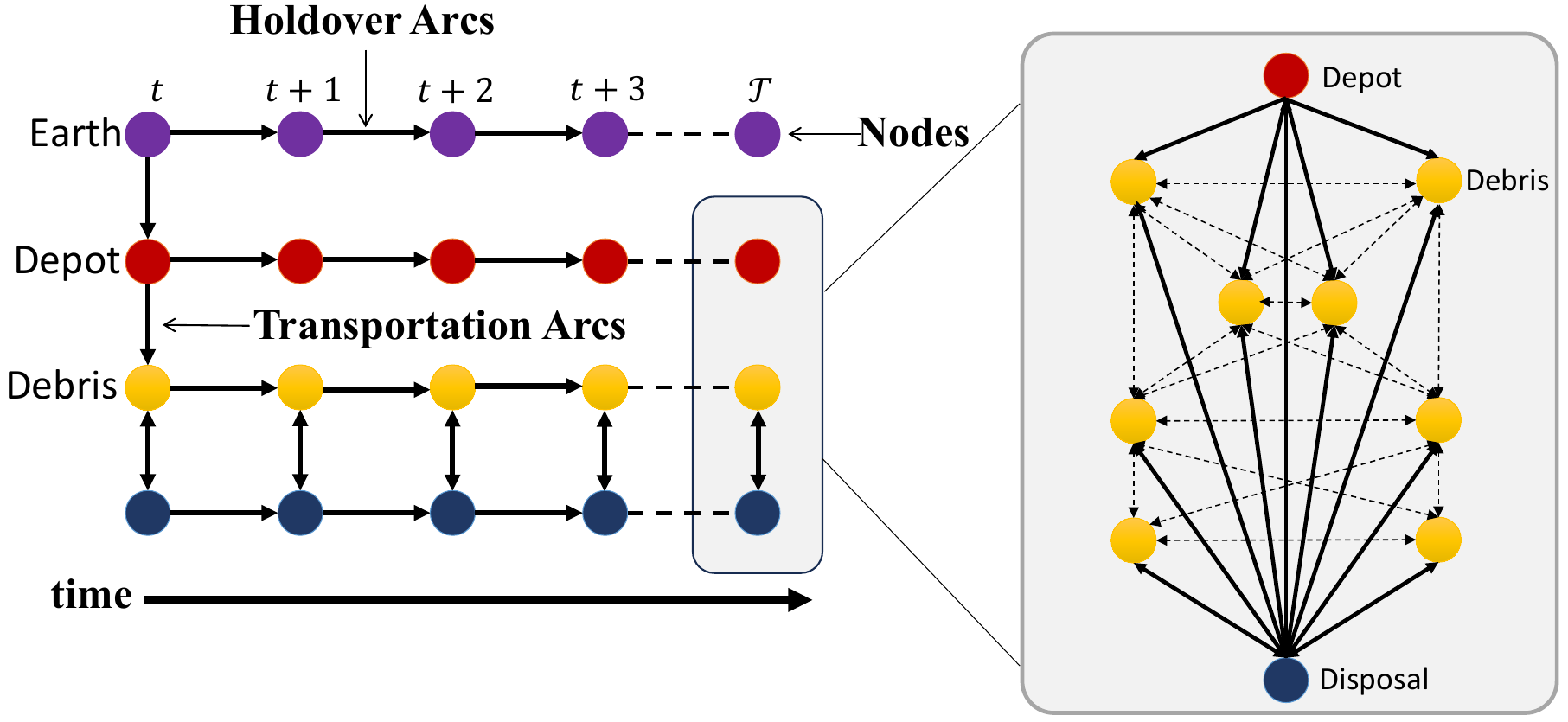}
    \caption{Logistics Network for Remediating Orbital Debris}
    \label{fig:networkdetail}
\end{figure}

Within our framework, we consider debris remediation operations as space logistics missions, modeling them as multi-commodity network flow problems. Consider a time-expanded network \(\mathcal{G}\), consisting of a set of nodes \(\mathcal{N}\) (index: \(\mathit{i,j}\)) and a set of arcs \(\mathcal{A}\). \rev{Since this is a time-expanded network, we define a set of time steps $\mathcal{T}$. To preserve computational efficiency, we assume that any orbital maneuver can be completed within a single time step. However, the real-time duration of each time step may vary.} Each arc has an index (\(\mathit{v,i,j}\)), corresponding to vehicle \(\mathit{v}\) flying from node \(\mathit{i}\) to node \(\mathit{j}\). Each vehicle \(\mathit{v}\) has variables for structural mass \(\mathit{s}_\mathit{v}\), commodity capacity \({\bm{e}}_\mathit{v}\), and the type of fuel it uses \(\mathit{f}_\mathit{v}\). The logistics decision variables are the commodity flows \(\ \bm{x}^\pm_{\mathit{vijt}} \) and the vehicle flows \(\ y^\pm_{\mathit{vijt}} \). These flows are split into inflows \(\ \bm{x}^-_{\mathit{vijt}} \)\(\mathit{/}\)\(\ y^-_{\mathit{vijt}} \) and outflows \(\ \bm{x}^+_{\mathit{vijt}} \)\(\mathit{/}\)\(\ y^+_{\mathit{vijt}} \). Outflows have cost coefficients \(\ \bm{c}^+_{\mathit{vijt}} \) for mission cost calculation. We assume that there are \(\mathit{p}\) types of commodities, making \(\ \bm{x}^\pm_{\mathit{vijt}} \) and \(\ \bm{c}^+_{\mathit{vijt}} \) vectors of size \(\mathit{p}\) \(\times\) 1, where each component corresponds to commodity flows and costs. The commodity flow variable \(\ \bm{x}^\pm_{\mathit{vijt}} \) is continuous for the propellant mass, and discrete for the debris objects being remediated. The vehicle flow variable  \(\ y^\pm_{\mathit{vijt}} \) is discrete. Each node has a demand vector \(\ \bm{d}_{\mathit{it}}\) which has positive values for mission supplies, and negative values for mission demands. Each arc from node \(\mathit{i}\) to \(\mathit{j}\) has a transit time \(\Delta\mathit{t}_\mathit{ij}\). The time-expanded network also considers time windows, which are specific time steps where commodity flows are permitted. The expression for the time window parameter is \(\mathit{W}_\mathit{ij} \subseteq [0, \mathit{T)}\), where \(\mathit{T}\) corresponds to the maximum mission time horizon. A detailed visualization of the logistics network can be seen in Fig. \ref{fig:networkdetail}.  

Each arc has flow transformation constraints and flow concurrency constraints. Flow transformation constraints show how commodities are consumed or converted during spaceflight, such as through fuel consumption or debris recycling. They are captured in the difference between outflows and inflows, and measured by transformation matrix \(\mathit{Q}_\mathit{vij}\). Flow concurrency constraints represent the flow upper bound defined by the spacecraft's commodity transportation capacity. It is captured by the concurrency matrix \(\mathit{H}_\mathit{vij}\), which splits commodities into categories of debris or propellant. A summary of these indices, variables, and parameters can be found in Table \ref{table:param}.

\begin{table}[h]
\centering
\caption{Definitions of Indices, Variables, and Parameters}
\label{table:param}
\begin{tabularx}{\linewidth}{ l X }

\hline
\hline
 Name & \multicolumn{1}{c}{Definition}  \\ 
 \hline
&\multicolumn{1}{c}{\textit{Index}}\\  
\hline
 \(\mathit{v}\)&Spacecraft index \\
 \(\mathit{i,j}\)&Node \\
 \(\mathit{t}\)&Time step \\
 \(\mathit{p}\)&Commodity index \\
 \(\mathit{l}\)&Concurrency index \\
 \hline
&\multicolumn{1}{c}{\textit{Variables}}\\
\hline
\(\bm{x}^\pm_{\mathit{vijt}} \)& Commodity outflows/inflows. Commodities in \(\ \bm{x}^\pm_{\mathit{vijt}} \) are considered as continuous or integer variables based on the commodity type. (\(\mathit{p}\) \(\times\) 1)\\
\(y^\pm_{\mathit{vijt}} \)& Number of spacecraft \(\mathit{v}\) flying along arc \(\mathit{i}\) to \(\mathit{j}\). Integer variable (scalar)\\
\(\mathit{s}_v\)& Structure mass of spacecraft \(\mathit{v}\). Continuous variable. (scalar)\\
\(\bm{e}_v\)& Spacecraft design parameters, including payload capacity. Continuous variables. (\(\mathit{l}\) \(\times\) 1)\\
\(\mathit{f}_v\)& spacecraft fuel type. Integer variable. (scalar)\\
\hline
&\multicolumn{1}{c}{\textit{Parameters}}\\
\hline
\(\bm{c}^+_\mathit{vijt}\)& Commodity cost coefficient. (\(\mathit{p}\) \(\times\) 1)\\
\(c'^+_\mathit{vijt}\)& Spacecraft cost coefficient. (scalar)\\
\(\bm{d}_{\mathit{it}}\)& Demands or supplies of different commodities at each node. (\(\mathit{p}\) \(\times\) 1)\\
\(d'_\mathit{vit}\)& Demand or supply of spacecraft \(\mathit{v}\) at each node. (scalar)\\
\(\mathit{Q}_\mathit{vij}\)& Commodity transformation matrix. (\((\mathit{p}+1) \times (\mathit{p}+1))\)\\
\(\mathit{H}_\mathit{vij}\)&  Concurrency constraint matrix. \((l \times p)\)\\
\(\mathit{W}_\mathit{ij}\)& Time window vector. (1 \(\times\) \(\mathit{n}\), where \(\mathit{n}\) is the number of time windows.)\\
\hline
\hline
\end{tabularx}
\end{table}

Based on these notations, the formulation for the time-expanded network can be expressed as:

\noindent Minimize 
\begin{equation}\label{objective}\mathcal{J} = \sum_{{t}\in \mathcal{T}}\sum_{{(v,i,j)}\in \mathcal{A}}(\bm{c}^\mathit{+}_{\mathit{vijt}}{^{T}{\bm{x}^\mathit{+}_\mathit{vijt}}} + {c}'^+_{vijt}\mathit{s}_{\mathit{v}}y^+_{\mathit{vijt}})\end{equation}

\noindent Subject to:
\begin{equation}\label{massbal1}\sum_{(v,j):(v,i,j)\in \mathcal{A}} \bm{x}^\mathit{+}_\mathit{vijt} -\sum_{(v,j):(v,j,i)\in \mathcal{A}}\bm{x}^\mathit{-}_\mathit{vji(t-\Delta\mathit{t}_\mathit{ji})} \leq \bm{d}_{\mathit{it}}\quad \forall\ i\ \in\ \mathcal{N}\quad \forall\ t\ \in\ \mathcal{T}  \end{equation}


\begin{equation}\label{massbal2}\sum_{j:(v,i,j)\in \mathcal{A}} y^\mathit{+}_\mathit{vijt} -\sum_{j:(v,j,i)\in \mathcal{A}}y^\mathit{-}_\mathit{vji(t-\Delta\mathit{t}_\mathit{ji})} \leq d'_{\mathit{vit}} \quad \forall\ v\ \in\ \mathcal{V}\quad \forall\ i\ \in\ \mathcal{N}\quad \forall\ t\ \in\ \mathcal{T}  \end{equation}


\begin{equation}\label{flowtrans}
\begin{bmatrix}
\bm{x}^-_{vijt} \\
{s}_{v}y^-_{vijt}
\end{bmatrix}=\mathit{Q}_\mathit{vij}\begin{bmatrix}
\bm{x}^+_{vijt} \\
{s}_{v}y^+_{vijt}\end{bmatrix}\quad \forall\ (v,i,j)\ \in\ \mathcal{A}\quad \forall\ t\ \in\ \mathcal{T}
\end{equation}

\begin{equation}\label{flowcon}
    \mathit{H}_\mathit{vij}x^\mathit{+}_\mathit{vijt}\leq {\bm{e}}_{v}y^+_{vijt}\quad \forall\ (v,i,j)\ \in\ \mathcal{A}\quad \forall\ t\ \in\ \mathcal{T}
\end{equation}

\begin{equation}\label{flowbou}
    \begin{cases}
\bm{x}^\pm_{\mathit{vijt}} \geq \bm{0}_{p \times1}\ \textrm{if}\ t\ \in\ \mathit{W}_\mathit{ij}  \\
\bm{x}^\pm_{\mathit{vijt}}= \bm{0}_{p \times1}\ \textrm{otherwise}
    \end{cases}\quad \forall\ (v,i,j)\ \in\ \mathcal{A}\quad \forall\ t\ \in\ \mathcal{T}
\end{equation}

\begin{equation}\label{routecon}
    \sum_{(v,j):(v,j,i)\in\mathcal{A}}y^{-}_{vji(t-\Delta{t}_{ji})}\leq 1\quad \forall\ i \in \mathcal{N}_d\quad \forall\ t\ \in\ \mathcal{T}
\end{equation}

\begin{align*}\label{vardefine1}\nonumber
    \bm{x}^\pm_{\mathit{vijt}}= {\begin{bmatrix}
    x_{1}\\
    x_{2}\\
    \vdots\\
    x_{p}\\
    \end{bmatrix}_{vijt}^\pm}
    \text{where} \ x_1 \ \in\ \mathbb{Z}_{\geq 0},\ x_2...,x_p \ \in\ \mathbb{R}_{\geq 0} \quad
    \forall\ (v,i,j) \in& \mathcal{A} \quad \forall\ t \in \mathcal{T}   \\
    y^\pm_{\mathit{vijt}}\in \mathbb{Z}_{\geq 0}  \quad \forall\ (v,i,j)\in \mathcal{A} \quad \forall\ t &\in \mathcal{T}\\
    s_v\in \mathbb{R}_{\geq 0},\quad \bm{e}_v\in \mathbb{R}_{\geq 0}^l,\quad f_v\in\mathbb{Z}_{\geq 0}\quad& \forall\ v \in \mathcal{V}
        \end{align*}

Equation \ref{objective} is the objective function, minimizing the cost of the orbital debris remediation mission. Equations \ref{massbal1} and \ref{massbal2} are mass balance constraints, limiting the commodity flows to satisfy node demands. Equation \ref{flowtrans} is the flow transformation constraint, showing how commodities are consumed or converted during spaceflight. Equation \ref{flowcon} is the flow concurrency constraint, which forms the upper bound of commodity flow based on spacecraft capacity. We consider propellant and debris capacity to be the concurrency constraints, and they are indexed by \(\mathit{l}\). Equation \ref{flowbou} corresponds to the flow bounds permitted by the time windows \(\mathit{W}_\mathit{ij} \subseteq [0, \mathit{T)}\). Equation \ref{routecon} is the vehicle routing constraint. \rev{We designate the nodes corresponding to debris locations as \(\mathcal{N}_d\).} Equation \ref{routecon} ensures that each debris node is visited \rev{at most} once. \rev{The resulting optimization problem formulation is solved using Gurobi Optimizer version 12.0.0.}

We consider three orbital debris remediation methods for our analysis. Two key assumptions for remediation maneuvers are that all remediation servicers use high-thrust propulsion systems, and that all orbital maneuvers are completed via Hohmann transfer. A full explanation of assumptions for remediation maneuvers can be found in Section \ref{Results and Analysis}. The details for each debris remediation method are as follows:

1) \underline{Controlled reentry}: The debris is transported from its original orbit to a circular Earth orbit at an altitude of 50 km \rev{\cite{Colvin_2023}}. Then, a $\Delta v$ of around 0.5 m/s is needed to initiate reentry \cite{Bacon}. The remediation satellite would capture the debris via a reversible process that will also allow for easy release of the debris at the prescribed altitude. This method allows for more precise disposal within a target zone, making it easier to ensure that no infrastructure or people are harmed by the de-orbiting debris. 

2) \underline{Uncontrolled reentry}: The debris is transported from its original orbit to a circular Earth orbit at an altitude of 350 km \cite{Colvin_2023}. This method offers less precision than controlled reentry, but benefits from requiring less propulsive capacity due to the higher altitude. For both reentry methods, after a single debris is remediated, the servicer vehicle either travels back to the depot for refueling or continues on to the next debris.

3) \underline{Recycling}: The debris is transported from its original orbit to a recycling center. This recycling center can extract metals from debris, which can be used for on-orbit manufacturing or propellant for metal plasma thrusters \cite{Calnan}. We assume that our remediation servicer uses high-thrust propulsion systems, so the recycling products won't be used to fuel the servicer. Instead, we consider the recycled product to be additional benefit for the recycling remediation method. The amount of benefit is calculated based on the avoided transportation cost of launching those metals from Earth.

\subsection{Incentive Design}





\begin{figure}[h]
    \centering
    \includegraphics[width=\textwidth]{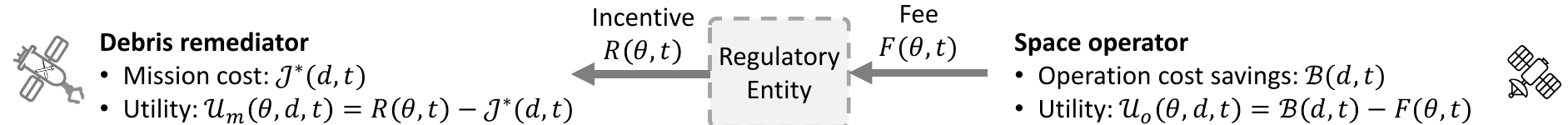}
    \caption{Overview of Incentive Mechanism}
    \label{fig:incentmech}
\end{figure}

Our framework uses a two-player cooperative game theory model to analyze the interactions between space operators and debris remediators. We define space operators as the owners of vehicles executing non-remediation space missions, whether for commercial or government purposes. We define debris remediators as the commercial entities who will design and execute orbital debris remediation missions. While multiple recent cost-benefit studies illuminate the economic value of orbital debris remediation \cite{Colvin_2023, Liou_2011, Liou_Johnson_Hill_2010}, the relationship established between space operators and debris remediators is one-sided. Debris remediators bear all the cost of removal, while space operators get to enjoy the benefits of an improved operating environment. We seek to identify incentive mechanisms to support more mutualistic relationships between operators and remediators. By using our logistics model in tandem with game theory, we can identify the role of governmental space agencies in mediating incentive mechanisms between space operators and debris remediators. Figure \ref{fig:incentmech} shows an overview of the incentive mechanism connecting the remediator and operator.

We define \( \mathcal{D}  \) to represent the set of all debris to be remediated in a given mission structure, and a variable \(\mathit{d}\) to represent the debris removal demand for the logistics model. The bargain between debris remediators and space operators contains an incentive and a fee. The incentive, \(R(\theta, t)\), is paid to the remediator for debris remediation. The fee, \(F(\theta, t)\), is collected from space operators for the service of remediating debris. This fee is a portion of the benefit that operators receive from remediated debris, \(\mathcal{B}(d,t)\). We define an incentive coefficient \(\theta\in[0,1]\) to specify how much fee needs to be collected from the operator's benefit \(F(\theta ,t)=\theta\cdot\mathcal{B}(d,t)\). We designate the utility of the debris remediator as \(\mathcal{U}_m(\theta,d,t)\). Based on the optimized mission cost \(\mathcal{J}^*\) from the space logistics model, we can write the utility of the remediator as
\begin{equation}\label{remutil}
    \mathcal{U}_m(\theta, {d}, t)  = R(\theta, t) - \mathcal{J}^*(d, t)
    \end{equation}

Similarly, we designate the utility of the space operator at time step \(t\) as \(\mathcal{U}_o(\theta, {d}, t)\). The utility of the operator is then written as
\begin{equation}\label{oputil}
    \mathcal{U}_o(\theta, d, t)  = \mathcal{B}(d, t) - F(\theta ,t)
    \end{equation}

Depending on the incentive mechanism, the fees collected \(F(\theta ,t)\) are not necessarily
the same as the incentive distributed \(R(\theta ,t)\). They can be asynchronous. For instance, an initial incentive can be used to cover the initial deployment cost of remediation servicers and facilities, with fees being collected gradually. In this case, fee collection and incentive distribution would be handled by a regulatory entity.

To quantitatively identify the best incentive mechanism, we use the Nash bargaining solution to perform function analysis on \(R(\theta,t)\) and \(F(\theta,t)\). The bargaining problem examines how players may share a jointly generated surplus. \rev{In order to formulate the problem, we must define a disagreement point, designated \(r\), which represents the strategies corresponding to the lowest utility expected by players if bargaining is unsuccessful. In this case, \(\mathcal{U}_m(\theta,d,t)\) and \(\mathcal{U}_o(\theta,d,t)\) are both 0.}

Nash proposed that the solution to the bargaining game is a utility vector that maximizes the Nash product \((\sum_{t\in \mathcal{T}}\mathcal{U}_m(\theta,d,t) - r_m)\cdot(\sum_{t\in \mathcal{T}}\mathcal{U}_o(\theta,d,t) - r_o)\) \cite{Nash_1950}. In our problem, the disagreement point is when no debris is remediated, so no cost is incurred by the remediator, and no benefit is gained for the operator. \rev{The fee and incentive would also be 0.} Thus, the disagreement point is \(r_m=r_o=0\). Substituting the expression of utilities into the Nash product, the Nash bargaining solution becomes:

\begin{equation}\label{argmax}
\underset{\theta}{\arg \max} \{\sum_{t\in \mathcal{T}}\mathcal{U}_m(\theta,{d}, t)\cdot\sum_{t\in \mathcal{T}}\mathcal{U}_o(\theta,{d}, t) | \sum_{t\in \mathcal{T}}\mathcal{U}_m(\theta,{d}, t) \geq 0,\sum_{t\in \mathcal{T}}\mathcal{U}_o(\theta,{d}, t) \geq 0\}
\end{equation}

We define the incentive \(R(\theta,t)\) and fee \(F(\theta,t)\) as mission state-dependent functions, where the debris remediator receives the incentive for each debris remediated and the space operator pays a portion of the benefit as the fee, \(R(\theta,t)=F(\theta,t)\). The utility of the space operator can be expressed as
\begin{equation}\label{thetaoputil}
    \mathcal{U}_o(\theta,d,t)=(1-\theta)\cdot\mathcal{B}(d,t)
\end{equation}
\noindent which is always non-negative. Similarly, the utility of the remediator can be expressed as 
\begin{equation}\label{utilsub}
    \mathcal{U}_m(\theta,d,t)=\theta\cdot\mathcal{B}(d,t)-\mathcal{J}^*(d,t)
\end{equation}
\noindent We designate the Nash product between utility expressions as $\mathcal{Z}(\theta)$. The complete substitution of these utility expressions yields the following optimization problem:


\noindent Maximize
\begin{equation}\label{nashbargain}
    \mathcal{Z}(\theta) =\sum_{t\in\mathcal{T}}(\theta\cdot\mathcal{B}(d,t)-\mathcal{J}^*(d,t))\cdot \sum_{t\in\mathcal{T}}((1-\theta)\cdot\mathcal{B}(d,t)) 
\end{equation}
Subject to: 
\begin{equation}\begin{split}\label{nashconstraints}
    \sum_{t\in\mathcal{T}}(\theta\cdot\mathcal{B}(d,t)-\mathcal{J}^*(d,t)) \geq 0\\
    \sum_{t\in\mathcal{T}}((1-\theta)\cdot\mathcal{B}(d,t)) \geq 0
\end{split}\end{equation}
\begin{equation*}\label{variabledefine2}
    \theta\ \in [0,1]
\end{equation*}

\section{Results and Analysis}\label{Results and Analysis}

\subsection{Benefit Analysis}

\begin{figure}[h]
    \centering
        \includegraphics[width=.7\textwidth]{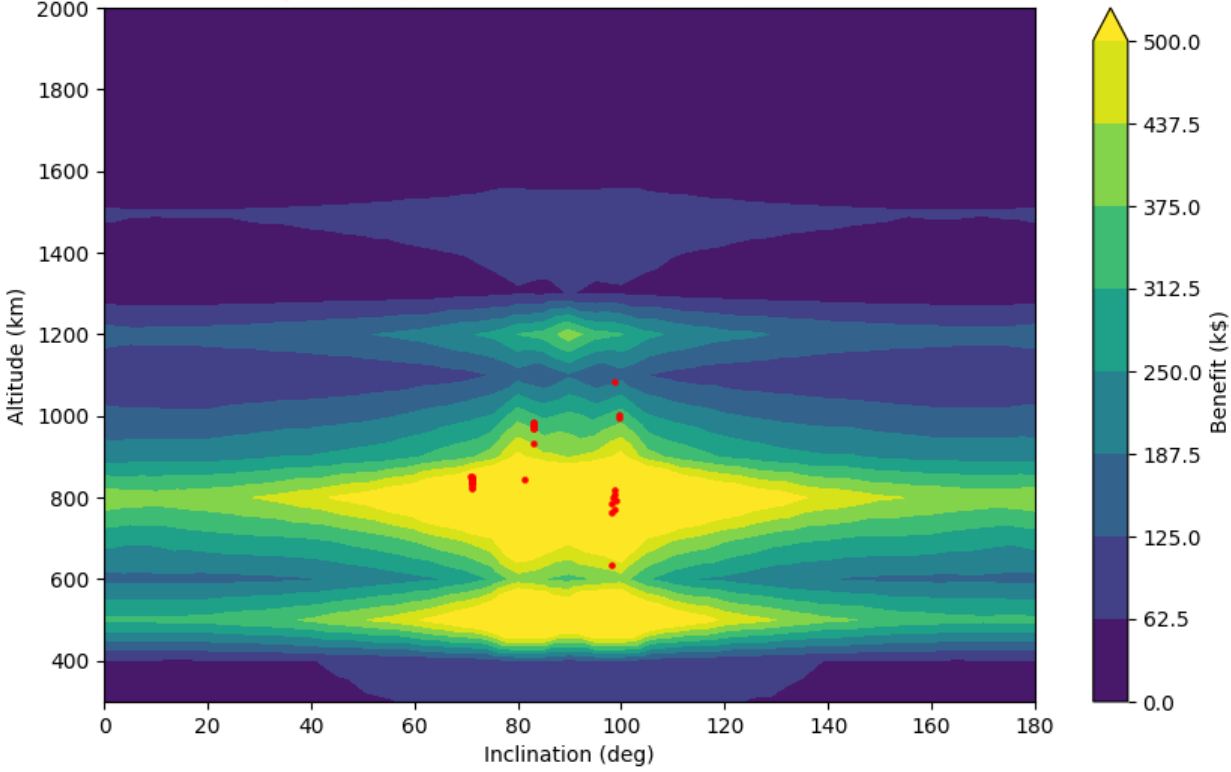}
        \caption {\rev{Yearly Benefit of Large Debris ($\geq$ 10 cm) Removal (Red Dots for Top 50 Most Concerning Debris)}}
    \label{fig:YearlyBig}
\end{figure}

To evaluate the yearly benefit of removing the top 50 most concerning debris objects \rev{listed by Ref \cite{McKnight2021}}, we consider the large debris itself and the potential small debris that it may generate from catastrophic collisions. \rev{We define large debris as having a characteristic length greater than or equal to 10 cm.}  Figure \ref{fig:YearlyBig} displays the total yearly benefit of removing the top 50 most concerning debris objects in LEO, including the additional benefit from preventing small debris generated through catastrophic collisions with the top 50 most concerning debris objects. The greatest benefit is localized around altitudes of 500 km and 800 km and between inclinations of 80\degree and 100\degree.

\subsection{Cost Analysis}

To demonstrate the effectiveness of the proposed commercialization framework, we utilize it to design a mission remediating the top 50 most concerning debris objects in LEO, per McKnight et al \cite{McKnight2021}. A table containing the relevant information about the top 50 can be found in Appendix \rev{Table \ref{table:A1}}, where the debris is ranked for remediation in order from most to least concerning. The mission cost \(\mathcal{J}\) and benefit \(\mathcal{B}\) are calculated as functions of the amount of debris being removed \(d\). \rev{The debris are indexed in order from most to least concerning. For our space logistics planning, we always remove the first $d$ most concerning debris.} We assume the spacecraft propellant capacity is ten times the spacecraft structural mass for debris remediation, corresponding with a propellant mass ratio of 0.9 \cite{oyedeko2021modelling,holt2009propellant}. \rev{This propellant capacity ensures that the servicer can complete at least one round trip to remediate a debris and return to the depot for refueling.} \rev{We also assume that there is no operational cost for recycling debris at the depot/recycling center.} Table \ref{table:missparam}~\cite{Colvin_2023,McKnight2021,leonard2023viability,RePEc:ebl:ecbull:eb-22-00204,Chen_Ho_2018} shows all assumptions for mission parameters and operation costs. 


\begin{table}[h]
\centering
\caption{Mission Scenario Assumptions}
\label{table:missparam}
\begin{tabular}{ l c }
\hline
\hline
\multicolumn{2}{c}{Mission parameter assumptions}\\
 \hline  
Parameters & Assumed Value  \\ 
 \hline  
 Spacecraft Propellant Capacity&10,000 kg \\
 Spacecraft Structure Mass &1,000 kg \cite{Colvin_2023} \\
 Propellant type&LH2/LOX \\
 Propellant \(I_{\textrm{sp}}\)&420 s \\
 Propellant \(\textrm{O}_2:\textrm{H}_2\) ratio &5.5 \\
 Average  Debris Mass& 5338 kg \cite{McKnight2021}\\
 Recyclable Debris Mass Percentage& 65\%\cite{leonard2023viability} \\
 \hline
\multicolumn{2}{c}{Operation cost assumptions}\\
\hline
 Rocket launch cost&\$1,500/kg \cite{RePEc:ebl:ecbull:eb-22-00204} \\
 Spacecraft Manufacturing &\$115M/spacecraft \cite{Colvin_2023} \\
 Spacecraft operation&\$45,000/flight \cite{Colvin_2023} \\
 LH2 price on Earth&\$5.94/kg \cite{Chen_Ho_2018} \\
 LO2 price on Earth&\$0.09/kg \cite{Chen_Ho_2018} \\
\hline
\hline
\end{tabular}
\end{table}

To account for the propellant consumption necessary for vehicles spaceflight when conducting debris removal missions, we compute the \(\Delta v\) values for maneuvers between nodes. Node IDs 1-50 refer to the 50 most concerning large debris objects in LEO. Node \(51_S\) refers to the location of a sun-synchronous on-orbit refueling depot used in the reentry remediation mission scenarios. We assume that the depot has been deployed prior to this mission structure, and do not account for its cost. The position of the depot was selected based on the average altitude of the 50 debris to be remediated. Node \(51_A\) refers to the location for a refueling depot/recycling center based on the average inclination of all top 50 debris objects, which is used in the recycling remediation mission scenario.  Nodes 52 and 53 correspond to uncontrolled reentry and controlled reentry disposal altitudes, respectively.

\begin{figure}[h]
    \centering
    \includegraphics[width = \textwidth]{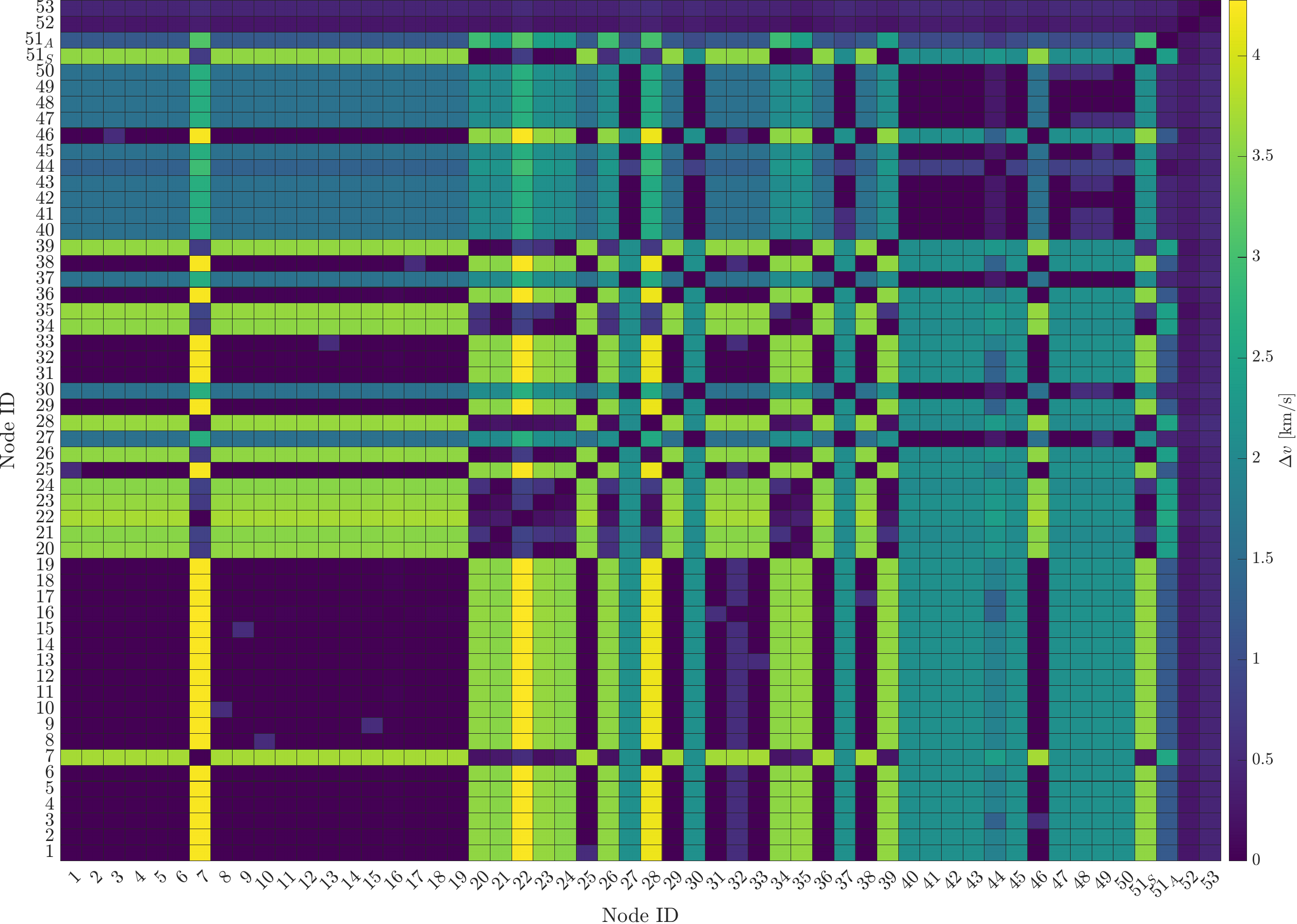}
    \caption{\rev{Heatmap of the Transfer Cost Between Orbit Nodes.}}
    \label{fig:HeatMap_c}
\end{figure}

Some key assumptions are made concerning the computation algorithm utilized to calculate the cost of orbital maneuvers between each orbital index. Firstly, all orbits are assumed to be circular. If an orbit is not circular (\textit{i.e.} the orbits of debris), the circular altitude is calculated as the average of the perigee and apogee altitudes. Secondly, it is assumed that the reentry orbits match the inclination of the opposing orbital index and that the argument of latitude is irrelevant for these orbits. Thirdly, it is assumed that any altitude change performed via a Hohmann transfer will account for any argument of latitude change (phasing) through a wait time before performing the altitude change, otherwise phasing is assumed to be 180 degrees. Fourthly, all plane changes occur at the higher altitude orbit, and phasing rendezvous occurs after the plane change since a plane change maneuver can only occur at a given argument of latitude where the planes coincide. Additionally, it is assumed that any maneuver that would involve a change in the right ascension of the ascending node (RAAN) waits an adequate amount of time such that the RAAN between the two orbits is the same, removing the need to consider a maneuver to change the RAAN. Finally, all transfers are high thrust, wherein the possible transfers include altitude, inclination, and argument of latitude (phasing rendezvous). The various transfer algorithms can be found in Ref.~\cite{Vallado2013}, where the boundary conditions are the orbital characteristics of the origin and destination orbital indices. \rev{Additionally, the phasing rendezvous algorithm from Ref.~\cite{Vallado2013} assumes allowance of up to four revolutions of the initial and transfer orbit so as to balance the maneuver cost with the time to perform the maneuver, while additionally keeping the time to perform the maneuver within a duration of eight hours}. A heatmap of the cost to transfer between each orbit node ID is shown in Fig. \ref{fig:HeatMap_c} with higher costs in a darker shade and the cost is for a transfer from the ID on the x-axis to the ID on the y-axis.

\begin{figure}[htp]
    \centering
        \includegraphics[width=\textwidth]{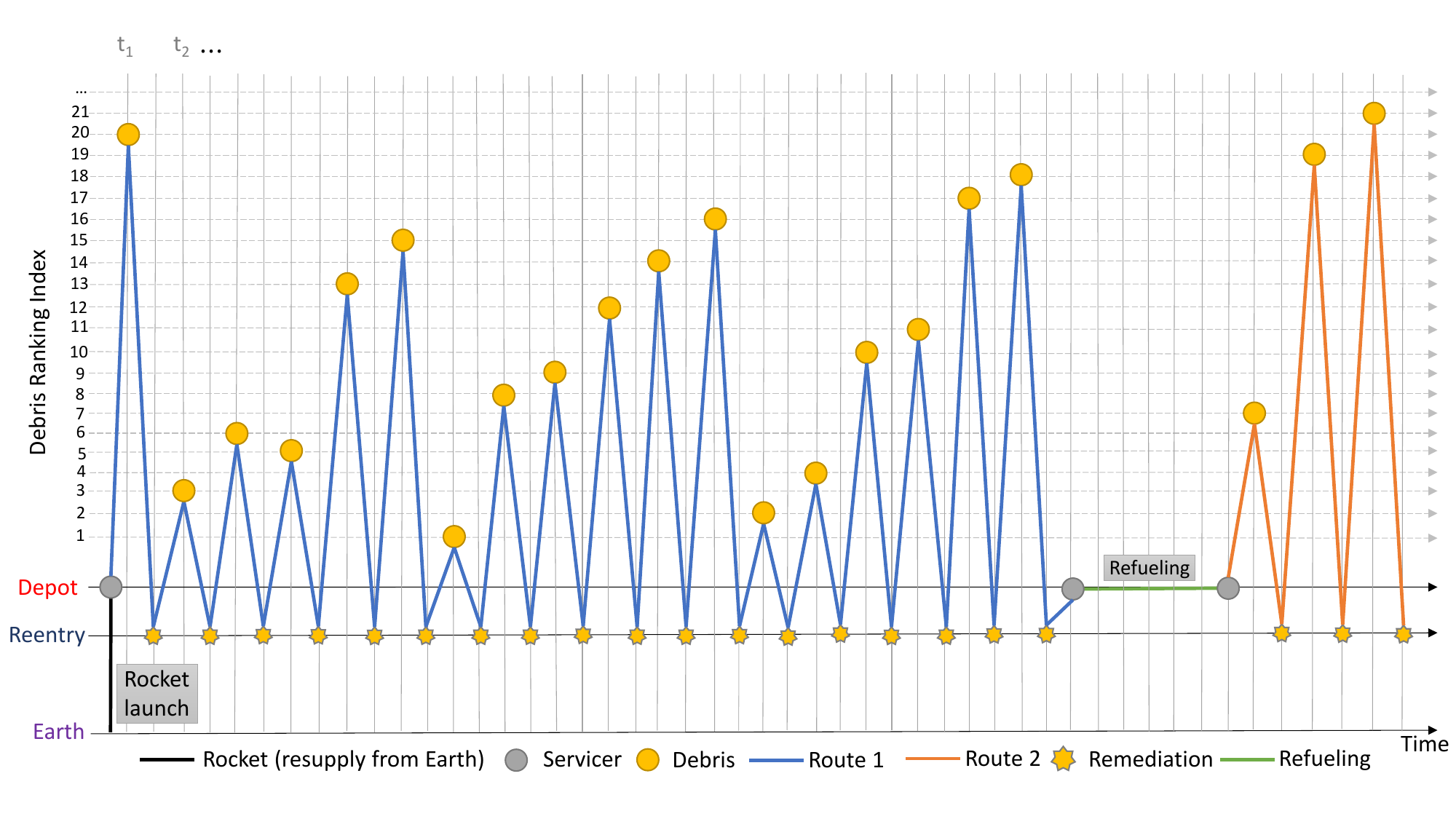}
        \caption{Logistics Planning Results for Remediating 21 Debris via Uncontrolled Reentry}
    \label{fig:UncontResult}
\end{figure}

The logistics planning results for remediating 21 debris via uncontrolled reentry is available in Fig. \ref{fig:UncontResult}. The servicer launches from Earth to the depot, and then from the depot proceeds to remediate 18 debris objects consecutively. After the 18th remediation, the servicer returns to the depot for refueling, and then leaves the depot to remediate the remaining three debris. Here it is visible that the order of remediation does not reflect the ranking of the most concerning debris, but reflects the optimal path to conserve fuel and minimize cost.

\begin{figure}[htp]
    \centering
        \includegraphics[width=\textwidth]{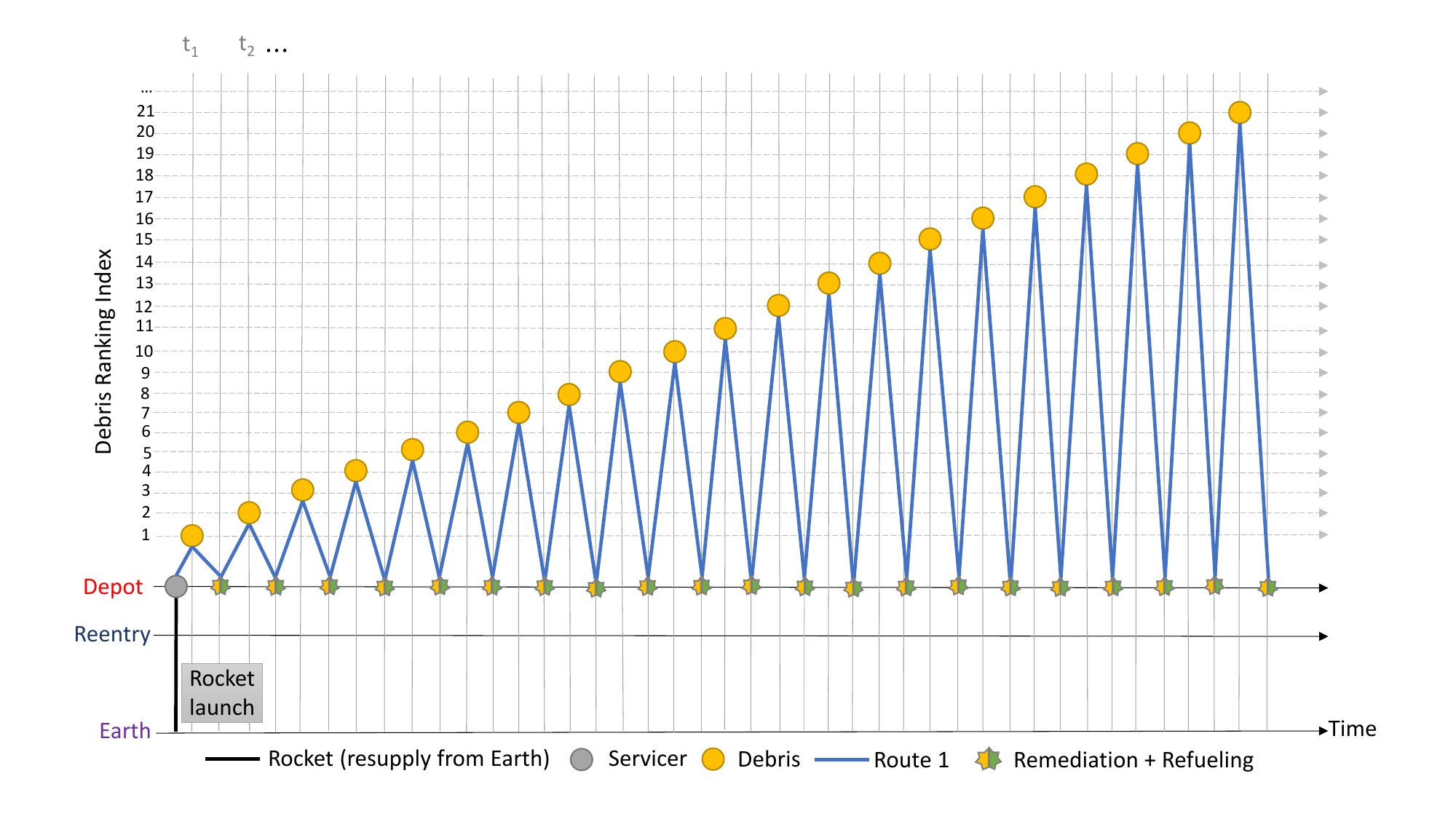}
        \caption{Logistics Planning Results for Remediating 21 Debris via Recycling}
    \label{fig:RecyclingResult}
\end{figure}

Logistics planning results for remediating 21 debris via recycling are seen in Fig. \ref{fig:RecyclingResult}. In contrast to the reentry remediation result, the order of the debris objects being remediated is irrelevant, as all remediation flights begin and end at the depot/recycling center. There is also only 1 route, since the servicer refuels each time it returns with debris. This method takes less time-steps compared to either reentry case, since there is one less location in the remediation route.

\begin{figure}[htp]
    \centering
        \includegraphics[width =.8\textwidth]{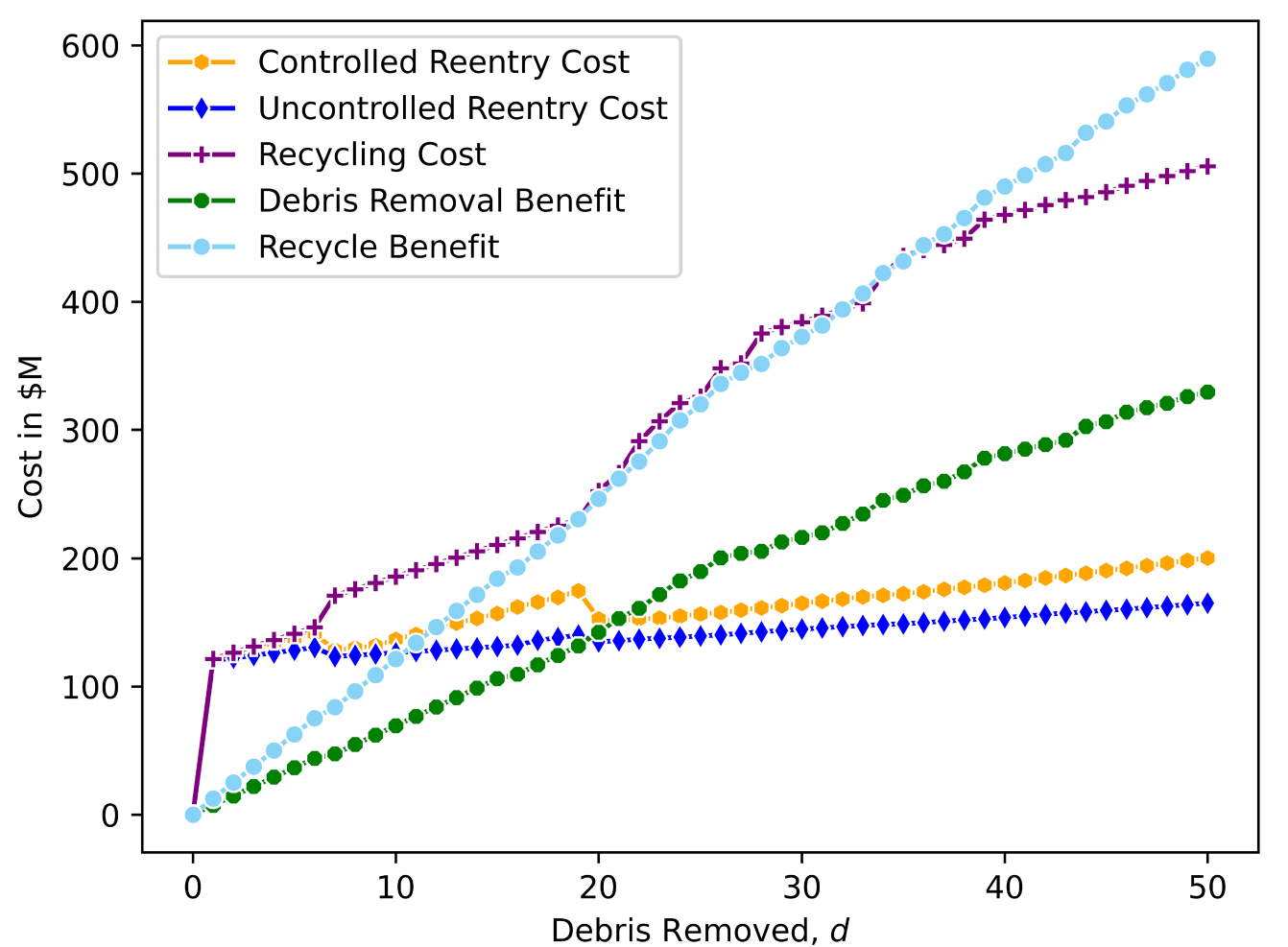}
        \caption{Mission Cost and Benefit to Remediate Up to \textit{d} Debris via Reentry and Recycling}
    \label{fig:DoubleCost}
\end{figure}

The mission cost for remediating \textit{d} debris via reentry methods is exhibited in Fig. \ref{fig:DoubleCost}. There are significant drops in the cost when removing 7 and 20 debris objects. This is related to the \(\Delta\mathit{v}\) cost for the flight between the depot and the first debris object to be remediated. Figure \ref{fig:HeatMap_c} shows that for most flights from the depot (Node \(51_S\)) to a debris object (Nodes 1-50), the \(\Delta\mathit{v}\) cost is very high, at nearly 4 km/s. There are exceptions to this trend when the 7th and 20th debris objects are included, as their \(\Delta\mathit{v}\) cost is significantly lower. To minimize the cost of the mission, the servicer will always start a remediation route with the lowest \(\Delta\mathit{v}\) cost. These dips in cost occur when there is a debris object that enables a much cheaper initial flight. The trends between results for controlled reentry and uncontrolled reentry are largely consistent. Controlled reentry is more expensive than uncontrolled reentry, which is expected since the lower disposal altitude necessitates a higher \(\Delta\mathit{v}\) cost. The addition of each new debris increases the total mission cost by about \$2 million, and the most expensive mission configuration has a cost of just over \$200 million.

The mission cost for remediating \textit{d} debris via recycling is also exhibited in Fig. \ref{fig:DoubleCost}. The cost for this remediation method continuously increases with each debris added, ultimately being over twice as expensive as the reentry remediation methods. This is because the recycling method requires leaving from and returning to the depot/recycling center for each debris remediation. As discussed for the reentry methods, this is the most expensive type of flight the servicer can make, since the depot/recycling center is a more precise location than disposing anywhere below a certain altitude. Having the depot/recycling center location \(51_A\) be based on an average debris inclination helps to minimize the total \(\Delta\mathit{v}\) cost, but the cost for recycling is still largely driven by the intense \(\Delta\mathit{v}\) demands.

Both graphs in Fig. \ref{fig:DoubleCost} include the benefit of removing or recycling debris, respectively. For the reentry methods, this number is calculated by multiplying the yearly benefit for removing \(d\) debris by 10, which we assume to be the number of debris-years remaining for an object if no remediation action is taken. At altitudes above 800 km, the natural lifetime of debris objects can be decades or centuries \cite{Kessler_Cour‐Palais_1978}. By considering 10 debris-years to be prevented by remediating these debris objects, we offer a conservative estimate of the cumulative benefit of remediating the designated debris. For the recycling remediation method, this benefit is further increased by the recycling of debris into metal for on-orbit manufacturing. Based on an average mass of 5338 kg, a recyclable debris mass percentage of 65\% \cite{leonard2023viability}, and the assumption that the benefit of recycling is tied to avoiding the cost of launching material into space, the additional benefit for recycling a debris object instead of just removing it is approximately \$5.2 million. For controlled reentry, the point where the benefit is greater than the cost occurs when 21 debris are removed. For uncontrolled reentry, it occurs when 20 debris are removed. For recycling, the benefit is briefly greater than cost when 32-34 debris are recycled, lower than the cost when 35 debris are recycled, then always greater than the cost when 36 or more debris are removed.  In each case, these points are significant because they correspond with the generated surplus that we explore with our incentive design. The bargaining deal has no basis until the benefit of removing debris outweighs the cost.

\subsection{Incentive Design}

\begin{figure}[htp]
    \centering
        \includegraphics[width=\textwidth]{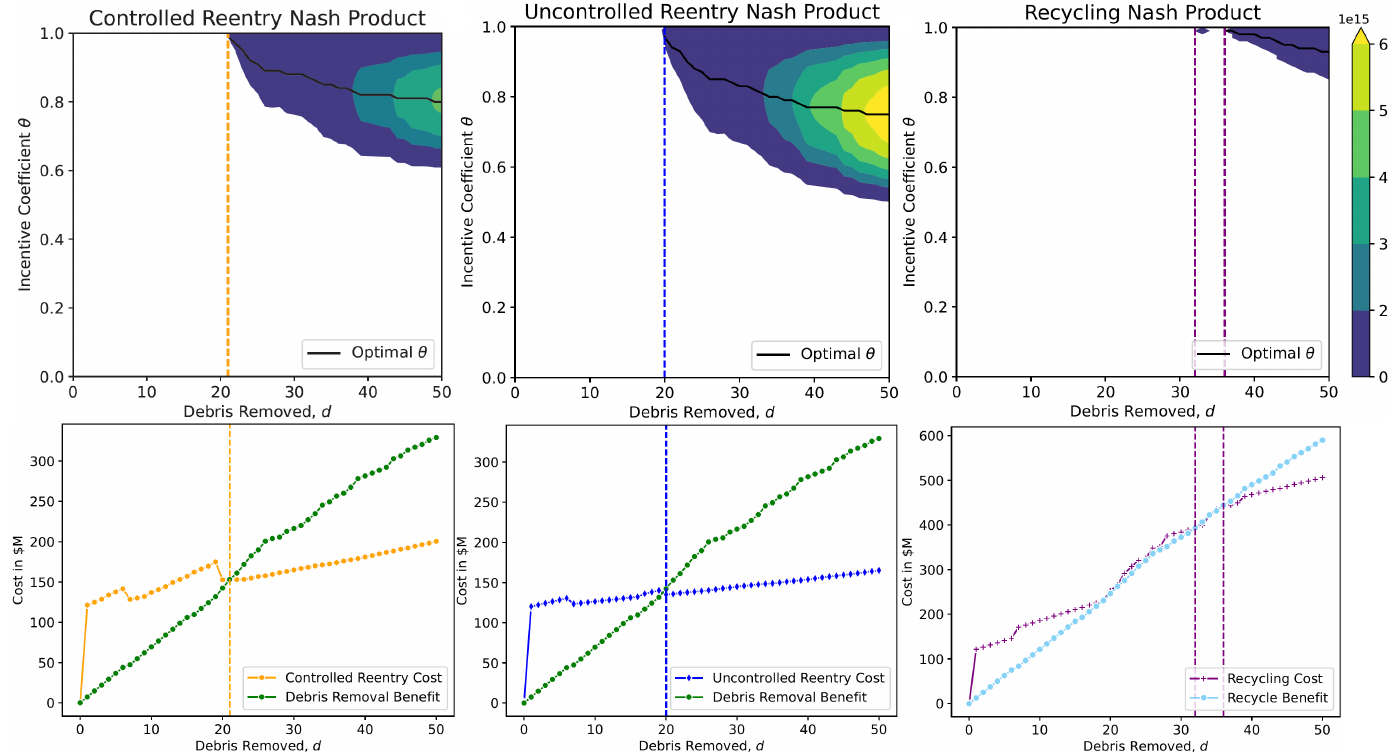}
        \caption{\rev{Nash Product for Controlled Reentry (left), Uncontrolled Reentry (center), and Recycling (right)}}
    \label{fig:UtilP}
\end{figure}

To analyze the commercial viability of remediating the top 50 most concerning orbital debris objects, we consider a bargaining problem between a space operator and a debris remediator. The solution to this bargaining problem is maximizing the product of each player's utility, also called the Nash product. The Nash products for remediation via controlled reentry, uncontrolled reentry, and recycling can be seen in Fig. \ref{fig:UtilP}. The dotted lines are used to illustrate that the bargaining policy space explicitly begins when the benefit is higher than the cost. As justified in the Cost Analysis section, the benefit used for the Nash product calculation was based on the assumption that each remediation prevents 10 debris-years.

The results of this analysis are promising for the commercial viability of remediating orbital debris. The positive values indicate that there is a generated surplus for remediating orbital debris, which allows the space for our incentive design. The region where the Nash product is highest represents the combination of incentive coefficients and debris removed that produce the greatest surplus for both players. The maximum Nash product is the same for both uncontrolled reentry and controlled reentry, but the region of that maximum benefit is significantly larger for uncontrolled reentry. This is expected based on our formulation. If the mission cost is lower, but the benefit remains the same, there will be an increase in the generated surplus, which is reflected in the Nash product. A general trend for both reentry remediation methods is that the more debris gets remediated, the less incentive must be implemented to reach the maximum Nash product. The recycling remediation method experiences a much smaller surplus than the other two methods, since the additional benefit from recycling still heavily contends with the higher mission cost. The Nash product reflects the noncontinuous surplus that was previously discussed in the cost analysis section.

\subsection{Sensitivity Analysis}

In order to measure how changing our assumptions would affect the results, we have conducted a sensitivity analysis for 3 major parameters: launch cost, recyclable mass, and debris-years prevented. These parameters are key assumptions within our framework, and by exploring them we provide insight into the range of success scenarios for commercial orbital debris remediation. 
\begin{figure}[htp]
    \centering
        \includegraphics[width=\textwidth]{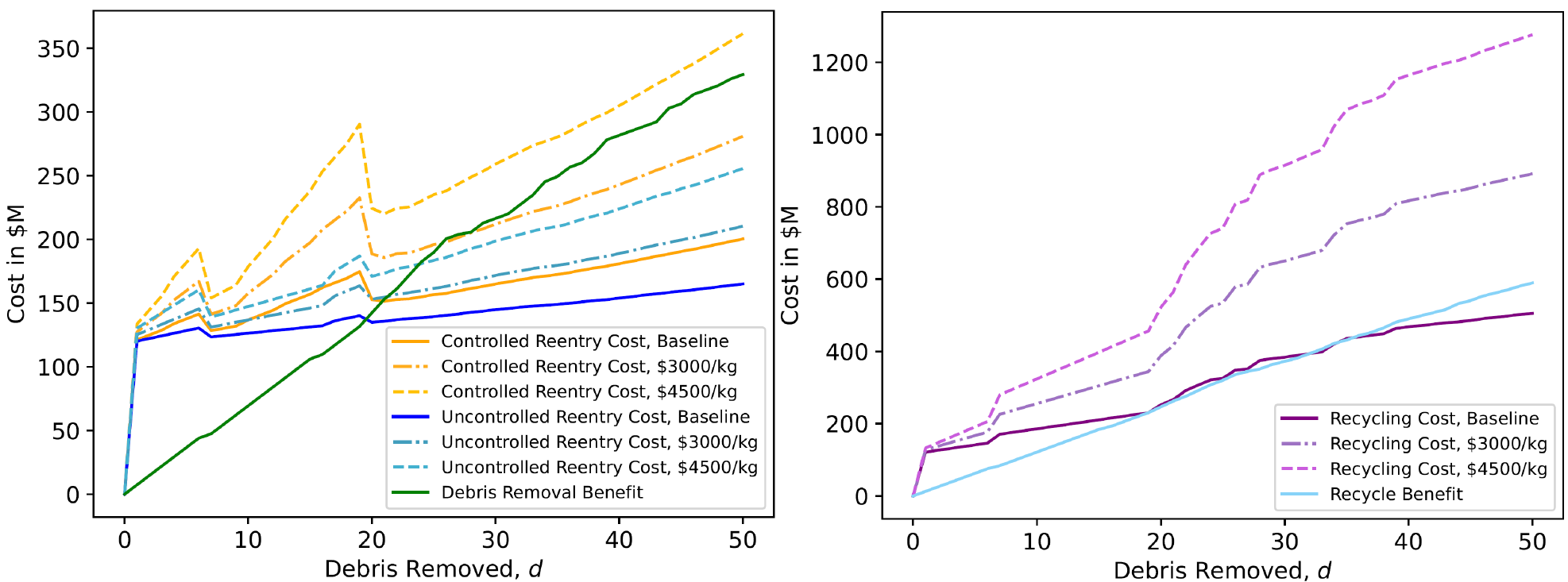}
        \caption{\rev{Launch Cost Sensitivity Analysis for Reentry Remediation (left) and Recycling Remediation (right)}}
    \label{fig:LCSensAn}
\end{figure}

First, we examine the effect of increasing launch cost from \$1,500/kg baseline to \$3,000/kg and \$4,500/kg, as seen in Fig. \ref{fig:LCSensAn}. \rev{The launch cost affects the launch of the servicer and all propellant.} For the reentry methods, the beginning of the surplus area where the benefit is greater than the cost, is predictably shifted forward as the launch cost increases. In the case of controlled reentry, the highest launch cost of \$4,500/kg completely prevents any surplus from being generated, as the mission becomes too costly. For recycling, the increases in launch cost both resulted in no surplus being generated, which was expected based on how limited the surplus area was in the nominal case. These results indicate that the launch cost is a significant factor in the viability of commercial debris remediation, as fluctuation in launch cost could make entire methods of remediation unviable.

\begin{figure}[htp]
    \centering
        \includegraphics[width=.65\textwidth]{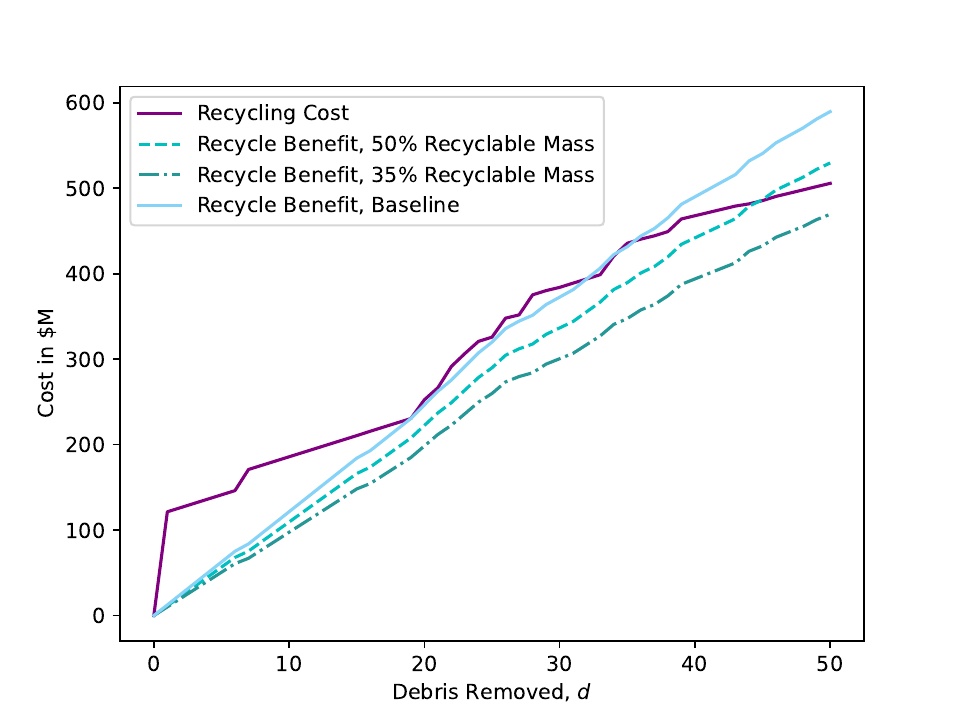}
        \caption{Recyclable Debris Mass Percentage Sensitivity Analysis}
    \label{fig:RMSensAn}
\end{figure}

In Fig. \ref{fig:RMSensAn}, we consider how diminished recycling capability might affect our policy space results. Results indicate that by reducing recyclable mass percentage from the 65\% baseline to 50\%, there is still a small amount of surplus generated after recycling almost all of the top 50 debris. An even smaller percentage of 35\% makes this method unviable, as the benefit never surpasses the cost. This highlights the importance of the technical feasibility and success rate that is necessary for recycling remediation to be commercially viable. 

\begin{figure}[htp]
    \centering
        \includegraphics[width=\textwidth]{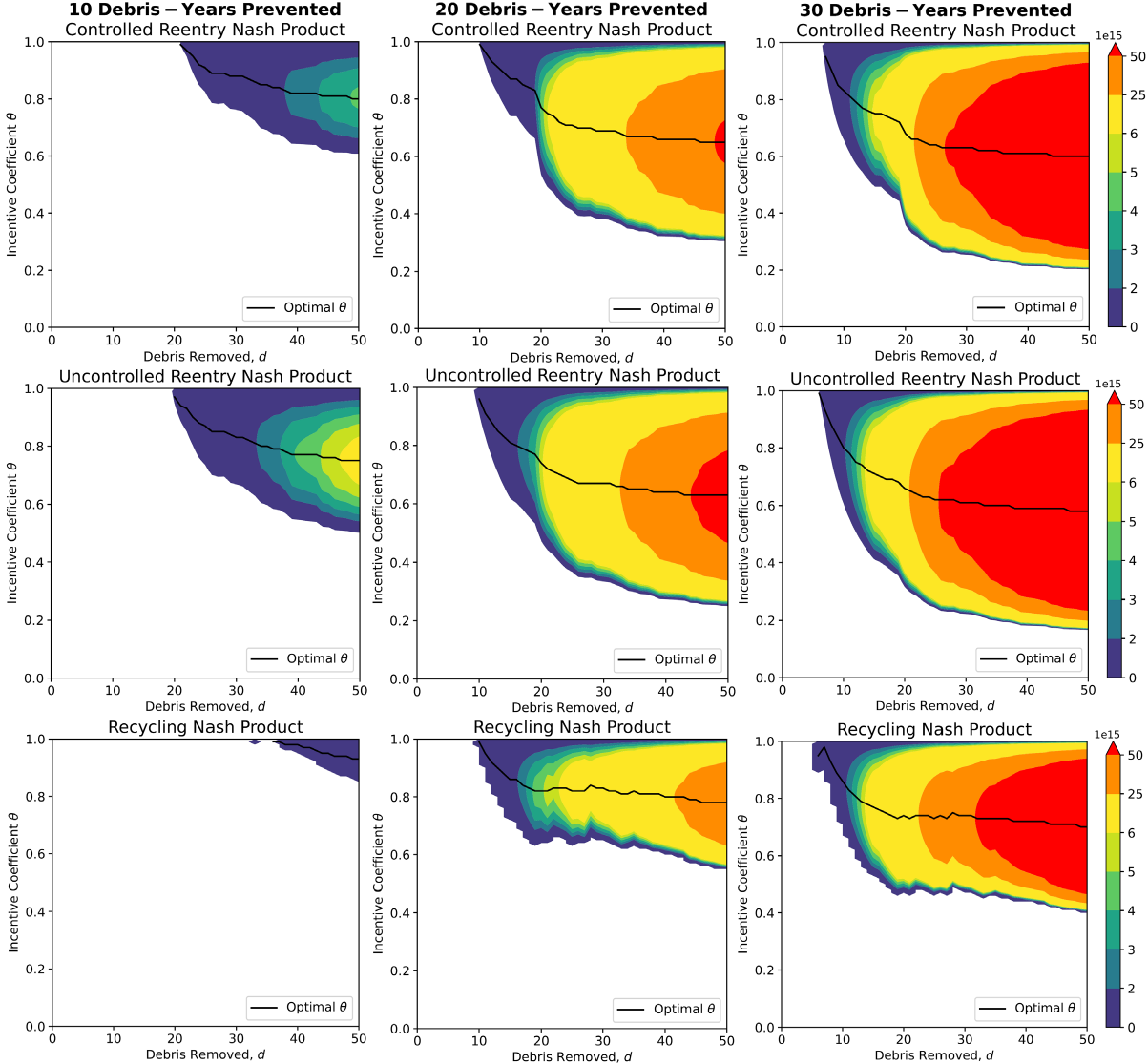}
        \caption{\rev{Debris-Years Prevented Sensitivity Analysis}}
    \label{fig:DYSensAn}
\end{figure}

Lastly, we consider how our assumption for debris-years prevented affects the bargaining space. While our nominal case only considers that each remediation prevent 10 years' worth of disruption from debris, it is likely that in reality that number is much higher. Figure \ref{fig:DYSensAn} examines how the bargaining space is affected by the debris-years prevented. These results indicate a substantial increase in shared surplus as more debris-years are prevented, which is expected since it substantially increases the benefit generated. The general trend of uncontrolled reentry being the most lucrative and recycling being the least lucrative is maintained. The Nash product being so heavily impacted by this parameter means that prevention of debris-years is a critical metric in measuring cost-effective, commercially viable remediation.

\section{Conclusion}\label{Conclusion}
This paper proposes a framework for the commercialization of orbital debris remediation. We use a space logistics network to optimize the cost of the remediation mission, a conjunction analysis model for calculating the benefit of remediating debris, and a game theory model to understand the ideal bargaining arrangement between space operators and debris remediators. In our framework, an incentive coefficient \(\theta\) is defined to measure the fee that must be collected from space operators and distributed to debris remediators, and a variable \(d\) is assigned to represent the number of debris objects to be remediated. Based on mission parameters, the bargaining problem is analyzed to find the configurations of \(\theta\) and \(d\) which generate the most preferable outcome for both players. To demonstrate the value of the framework, we evaluate a mission structure to remove the 50 most concerning debris objects in LEO. The results of our analysis indicate that there is a surplus generated from the remediation of orbital debris, and that there are various configurations that allow this surplus to be optimally shared by space operators and debris remediators. The future of this work can focus on capturing more of the complexities of mission and incentive design. These include reducing simplifying assumptions for cost estimations, using more precise estimates for debris-years prevented per object, and considering a wider variety of players in the incentive design.

\section*{Acknowledgments}
This material is based upon work supported by NASA under award No. 80NSSC24K0057. Any opinions, findings, and conclusions or recommendations expressed in this material are those of the authors and do not necessarily reflect the views of the National Aeronautics and Space Administration. Hang Woon Lee acknowledges additional partial support from NASA award No. 80NSSC23K1499.

\section*{Appendix}
\setcounter{table}{0}
\renewcommand{\thetable}{A\arabic{table}}
    
\begin{center}
\begin{longtable}[c]{l l l l l l }
\caption{Top 50 Most Concerning Debris (data extracted from \cite{McKnight2021})} \\
\hline
\hline
Ranking&SATNAME&APOGEE, km& PERIGEE, km&INCL., deg&MASS, kg\\
 \hline  
1&SL-16 R/B&848&837&71.0&9000\\   
2&SL-16 R/B&848&827&71.0&9000\\
3&SL-16 R/B&846&843&71.0&9000\\
4&SL-16 R/B&854&827&71.0&9000\\
5&SL-16 R/B&844&833&71.0&9000\\
6&SL-16 R/B&853&834&71.0&9000\\
7&SL-16 R/B&1006&986&99.5&9000\\
8&SL-16 R/B&852&831&71.0&9000\\
9&SL-16 R/B&844&835&71.0&9000\\
10&SL-16 R/B&845&838&71.0&9000\\
11&SL-16 R/B&846&823&71.0&9000\\
12&SL-16 R/B&845&841&71.0&9000\\
13&SL-16 R/B&844&840&71.0&9000\\
14&SL-16 R/B&850&823&71.0&9000\\
15&SL-16 R/B&848&831&71.0&9000\\
16&SL-16 R/B&863&839&70.8&9000\\
17&SL-16 R/B&848&842&71.0&9000\\
18&SL-16 R/B&841&831&71.0&9000\\
19&SL-16 R/B&842&814&71.0&9000\\
20&SL-16 R/B&813&801&98.6&9000\\
21&ENVISAT&766&764&98.1&7800\\
22&METEOR 3 M&1013&994&99.6&2500\\
23&ADEOS&794&793&98.9&3560\\
24&H-2A R/B&836&734&98.2&3000\\
25&SL-12 R/B(2)&847&838&71.0&2440\\
26&CZ-2D R/B&846&791&98.7&4000\\
27&SL-8 R/B&9956&966&82.9&1435\\
28&H-2 R/B&1306&860&98.7&2700\\
29&COSMOS 2322&854&842&71.0&3250\\
30&SL-8 R/B&992&961&82.9&1435\\
31&COSMOS 2406&863&844&71.0&3250\\
32&COSMOS 2278&852&841&71.1&3250\\
33&COSMOS 1943&851&833&71.0&3250\\
34&ADEOS 2&801&800&98.5&3680\\
35&SL-16 R/B&645&622&98.2&9000\\
36&SL-12 R/B(2)&848&794&71.1&2440\\
37&SL-8 R/B&989&957&83.0&1435\\
38&COSMOS 1844&866&824&71.0&3250\\
39&ARIANE 5 R/B&796&748&98.6&2575\\
40&SL-8 R/B&981&955&82.9&1435\\
41&SL-8 R/B&992&950&82.9&1435\\
42&SL-8 R/B&1001&970&82.9&1435\\
43&SL-8 R/B&996&954&82.9&1435\\
44&SL-3 R/B&896&791&81.3&1100\\
45&SL-8 R/B&999&969&82.9&1435\\
46&COSMOS 2082&856&833&71.0&3250\\
47&SL-8 R/B&996&953&82.9&1435\\
48&SL-8 R/B&988&966&83.0&1435\\
49&COSMOS 1275&1014&954&83.0&800\\
50&SL-8 R/B&996&953&82.9&1435\\
\hline
\hline
\label{table:A1}
\end{longtable}
\end{center}

\clearpage
\newpage
\bibliography{references}

\end{document}